\documentclass[sigconf]{acmart}

\usepackage{soul,xcolor}

\usepackage{balance}       
\usepackage{graphics}      
\usepackage[T1]{fontenc}   
\usepackage{algorithm} 
\usepackage{algpseudocode}
\usepackage{mathptmx}
\usepackage{color}
\usepackage{booktabs}
\usepackage{textcomp}
\usepackage{microtype}        
\usepackage{ccicons}          
\usepackage{todonotes}
\usepackage{subcaption}
\usepackage{cleveref}

\def\BibTeX{{\rm B\kern-.05em{\sc i\kern-.025em b}\kern-.08emT\kern-.1667em\lower.7ex\hbox{E}\kern-.125emX}}

\copyrightyear{2019}
\acmYear{2019}
\setcopyright{acmlicensed}
\acmConference[Intelligent Information Feed '19]{Intelligent Information Feed workshop '19}{August 05, 2019}{Anchorage, Alaska, USA}
\acmBooktitle{Intelligent Information Feed workshop '19, August 05, 2019, Anchorage, Alaska, USA}
\acmPrice{15.00}
\acmDOI{xx.xxx/xx.xxx}
\acmISBN{xxx-x-xxxx/xx/xx}

\begin{document}

\setstcolor{red}

%
\title{Fairness and Diversity in the Recommendation and Ranking of Participatory Media Content}
\title[Fairness and Diversity in Recommendation Systems]{Fairness and Diversity in the Recommendation and Ranking of Participatory Media Content}
%

\author{Muskaan*}
\affiliation{%
 \institution{Indian Institute of Technology}
 \city{Delhi}
 \country{India}
}
\email{cs1150240@iitd.ac.in}

\author{Mehak Preet Dhaliwal*}
\affiliation{%
 \institution{Indian Institute of Technology}
 \city{Delhi}
 \country{India}
}
\email{cs1150238@iitd.ac.in}

\author{Aaditeshwar Seth}
\affiliation{%
 \institution{Indian Institute of Technology}
 \city{Delhi}
 \country{India}}
\email{aseth@cse.iitd.ac.in}

\thanks{*These authors contributed equally to the paper.}






%
\renewcommand{\shortauthors}{Mehak, Muskaan and A.Seth}

%
\begin{abstract}
Online participatory media platforms that enable one-to-many communication among users, see a significant amount of user generated content and consequently face a problem of being able to recommend a subset of this content to its users. We address the problem of recommending and ranking this content such that different viewpoints about a topic get exposure in a fair and diverse manner. We build our model in the context of a voice-based participatory media platform running in rural central India, for low-income and less-literate communities, that plays audio messages in a ranked list to users over a phone call and allows them to contribute their own messages. In this paper, we describe our model and evaluate it using call-logs from the platform, to compare the fairness and diversity performance of our model with the manual editorial processes currently being followed. Our models are generic and can be adapted and applied to other participatory media platforms as well.
\end{abstract}

\maketitle

\section{Introduction}

Participatory media systems such as for online social networking, discussion forums, blogs, content collaboration platforms, etc are characterized by discussions among members with views, arguments, and counter arguments about different topics actively put forth by the users. Such systems are known to face problems ranging from echo chambers that arise when sub-communities are formed with homogenized views and are closed to other opinions \cite{usblogosphere}, algorithmically created filter bubbles that strengthen echo chambers by picking up signals arising due to a polarization of viewpoints \cite{facebookfeed, trumpfakenews}, and even disengagement and hostility when users are faced with viewpoints that they disagree with \cite{facebookdisengagement, twitterhostility}. Prior work has looked at content recommendation solutions that specifically aim to reduce bias in the recommendations of blogs and comments on blogs by maximizing the completeness of views shown to a user on a given topic \cite{sethicwsm}. More recent work has looked at the problem of fairness in ranking for news articles shown to a user, to create a balance between ranking relevant content higher but at the same time ensuring that diverse viewpoints are represented as well \cite{elisa}. We build upon these research directions to come up with a content recommendation and ranking strategy especially suited for participatory media platforms where multiple users might be engaged in a discussion on a given topic but may have different preferences and views, while the platform providers may want to ensure a certain degree of fairness and diversity so that even minority viewpoints get attention.

We model a given topic as being comprised of multiple sub-topics, which we refer to as aspects. Different users may have different preferences towards these aspects, or even different sentiments towards these aspects. A recommendation system may be able to learn these preferences and accordingly recommend to the users only content contributed by other users which will be in line with their preferences. We want to intervene at this stage, and provide to the platform managers a framework whereby they can specify different kinds of editorial policies. These policies could specify, for example, either to just recommend content in line with the user preferences, or to override it to the extent of giving some minimum amount of exposure to each aspect or each sentiment of each aspect, or to give equal exposure to various aspects, etc.
We consider that the platform is visited by different users who contribute to discussions on the given topic, and are exposed to content contributed by other users as specified by the editorial policies. Our framework however meets the editorial policy specifications not at a per-user level, but on an aggregate population level, ie. across all the users who visit the platform, our framework can ensure that each aspect gets some minimum exposure, or equal exposure, etc. To avoid trivial solutions where a randomly chosen user could be simply shown content from a single aspect, we add an additional constraint for diversity in the list of content recommended to a user. This overall setup which we term as short-term diversity with long-term fairness, recommends content so that each individual user gets a diverse listing whenever they access the platform, and across all the users who interact on a given topic the editorially specified fairness policies are also honoured.

This setting is grounded in a real-world requirement. We work in collaboration with a social enterprise in India, Gram Vaani, that operates a voice-based community media network in rural areas, aimed at enabling information access and sharing to less literate populations \cite{aparnaictd2016}. This platform, called Mobile Vaani (MV), runs using IVR (Interactive Voice Response) systems that can be accessed even via simple non-Internet based mobile phones. Users can call into the system and listen to audio messages left by other users, and leave their own messages. Voice messages recorded by the users are reviewed manually by a team of moderators, and the accepted messages are published on the IVR system for other users to hear. A wide range of topics are discussed, including local events and policies, government schemes, social norms such as early marriage and domestic violence, peer-to-peer facilitated agricultural question answering, etc. Since IVR systems can be unwieldy to have users select through key-presses the specific topics in which they are interested, a broadcast-radio inspired design is used on MV, where a programming chart is prepared to have the IVR rank content at the top for different topics at different pre-specified times of the day. For example, an agriculture slot may run on Mondays from 4-6pm, a health slot may run from 6-8pm, etc. Similarly, some important topics related to current events may be assigned their own slot in which users can listen and contribute their views. It is specifically for such slots on contentious topics that there is a need for the content recommendation and ranking model we have described above. Users on the platform may come from diverse backgrounds and ideologies, and may have preferences to listen to only certain aspects or sentiments about the topic, but the MV managers want to impose editorial policies so that fairness and diversity can be ensured too.

We use data from the platform for several health and nutrition related topics that were discussed actively \cite{jeevikaictd2019}. Although this is not as contentious as discussions on regional and national policies, the health data was sufficiently voluminous for algorithmic analysis since it was part of a funded project for the social enterprise, and showed a fair degree of variation in user preferences. This data therefore allows us to evaluate our algorithms. The social enterprise is now releasing a mobile application as well, which will make it easier for users to navigate to a specific topic, and hence the same algorithms can be applied on the mobile application to operate within topic-specific discussions. In the same way, the model can be adapted to other participatory media platforms as well where users contribute messages and view each others messages.

Recommending content that features views to which a user may be opposed, can result in disengagement or even hostility. We feel however that the character of the medium if shaped appropriately, can help guard against such problems. In the voice-based participatory media platform described here, the enterprise has reported several best practices they incorporated in their operations which have helped in developing norms for users to have mutual respect for one another and for diverse viewpoints \cite{responsibleoutcomes}. An emphasis throughout for users to carry a respectful tone in their contributions, and manual selection by the moderators of diverse viewpoints to be featured in prominence, has helped create a culture of debate on Mobile Vaani. Algorithms which can carry such an editorial policy forward, and can also be applied to other platforms, can therefore be important in creating a responsible media ecosystem where diversity is appreciated rather than denounced.

\begin{figure*}[h]
    \centering
    \includegraphics[width=\linewidth]{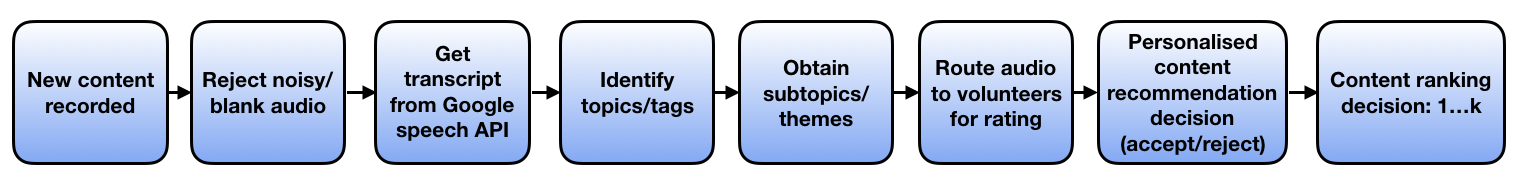}
    \caption{Content Pipeline: Stages of processing undergone by new items which are recorded by users}
    \label{fig:content_pipeline}
\end{figure*}


\section{Related work}

The issues of algorithmic fairness and algorithmic accountability have seen a lot of attention lately \cite{cathyoneil}, and illustrate how intentions of the designers and managers of technology systems can get encoded in the operation of the systems, consciously or unconsciously. Our work is an attempt to allow the platform managers of participatory media systems to define their editorial policies transparently, and provide an algorithmic means to impose these policies on the platform. 

Our problem relates most closely to the issue of fairness in ranking of documents. Ranking has been studied in the context of search engines where concerns have been raised on the potential for search results to influence user opinions about political candidates, gender and racial biases, etc \cite{10, 11, 12}, and has led to debates about search neutrality \cite{grimmelmann2010some, crane2012search}. Several research works \cite{elisa, fr1, fr2} have proposed a greedy algorithm for fairness in such search rankings, which ensures that no class of content is over-represented in the ranking based on constraints that can be specified as an editorial policy. The context in which we face this problem is similar. Our goal is to ensure fairness in the recommendation of audio messages on a voice-based participatory media platform, Mobile Vaani, operating in rural areas in India \cite{aparnaictd2016}, so that no single ideology or viewpoint gets more prominence than others. Other similar voice-based discussion forums such as CGNet Swara \cite{cgnet} and Avaaj Otalo \cite{avaaj} can also benefit from solving this issue.

Our algorithm builds upon the approach of Celis, et al \cite{elisa}. Their algorithm generates a single ranked list maximizing the utility while obeying class-specific fairness constraints. However, our requirement is to generate such ranked lists multiple times for various users visiting the MV platform, to ensure class-specific fairness across these multiple runs. Hence we extend the approach of Celis, et al by incorporating a concept of memory that remembers how much exposure has been assigned to content belonging to different classes until now, and takes it into account when producing the next ranked list of content. This setting of multiple ranked lists is also defined by Biega, et al \cite{fr3} who generate a series of rankings to assign exposure proportional to the relevance of the documents. Their work is complementary to ours and solves the problem through linear programming while we have designed a simple greedy approach.

Celis, et al also use the terms diversity and fairness interchangeably, whereas we distinguish between these terms. We define diversity as a short-term goal to ensure that in any ranked list, multiple classes are shown. Fairness is defined as a long-term goal, that across multiple lists accessed by users over time, no class is over-represented, or as defined by a fairness policy some minimum representation is given to all the classes. This is more similar to the definitions provided in \cite{pop} to an extent, where diversity is defined as the ``dissimilarity'' of items being recommended and fairness is defined as having a balance between items of different classes in the recommendation. Our problem is somewhat different from \cite{pop} though, since we only generate short lists of items at one time due to user browsing characteristics where only a few users go deep down a list of audio items, and hence we need to operate with multiple lists instead of ensuring diversity and fairness within a single list.

Celis, et al also assume that the relevance of items is known in advance as a utility measure for each item, and it serves as an input for their algorithm. Some works such as by Singh, et al \cite{singh2018fairness} combine the relevance and ranking problems. We choose to keep these problems separate from each other, and first using a simple machine learning classifier we generate recommendations for users based on their preferences that have been learned by the classifier, then we operate on these recommended items to generate ranked lists that ensure diversity and fairness. This allows us to solve the recommendation problem separately, because it is highly dependent on the amount of data that we have. We in fact do not have enough data per-user to be able to build personalized recommendations for each user, rather as we explain later we identify groups of users similar to one another and build a recommendation system customized to each group. We do not focus much on the recommendation problem in this paper, we only build a basic functional solution and instead investigate the diversity and fairness properties of our ranking algorithm in detail.

The recommendation step in our work can be substantially enhanced though, to prevent classification biases that might arise due to biases in the data. Work such as \cite{fc1, fc2} ensure different forms of statistical fairness towards individuals or groups, where protected variables may be correlated with other variables and thereby the classification may produce biased results on the protected classes. Such methods can be incorporated into our work in the future, to impose editorial policies in the recommendation step as well.

\section{Algorithmic model and data preparation}
As described earlier, different programming slots are assigned to different topics. Decisions of which topics should be assigned their own slot, and when should the slot be activated, are made by the content operations team of the social enterprise. This internally involves a consultative process between the content managers and community reporters in the field, a trend analysis of which topics are getting more attention to deserve their own slot, whether or not a prime-time slot can be allocated to a topic (Figure \ref{fig:users_per_hour} shows the average number of users who call at different hours in the day over three months), etc. Once slots have been assigned, and new content items are contributed by users, the moderation team reviews them and marks relevant aspects for them. Currently this is done manually, and efforts are underway to automate this as much as possible using speech2text technologies. For context, the pipeline that any new item needs to follow is shown in figure \ref{fig:content_pipeline}. An accept/reject classifier first determines whether the audio quality of the item is good enough for acceptance, accepted audio items are passed on to a speech engine that returns the transcript, the transcript is then passed on to other classifiers to determine the topic and aspects within the topic. This information is then used by a personalized content recommendation classifier that we are building, to decide whether to recommend the content to a given user or not based on the learned preferences about the user. Finally, a ranking algorithm that ensures short-term diversity and long-term fairness decides how to rank content predicted to be of interest to the user. Our focus in this paper is on the ranking algorithm rather than on content recommendation; we use a simple recommendation model currently and treat it like a blackbox for the purpose of evaluating the ranking algorithm.

Within the IVR, the audio items for any topic slot are presented in a list, one after the other. Users can press a button to skip to the next item, and at any point they can record their own item. Other features are also available such as liking a message, forwarding it, etc. Our algorithm needs to generate this list of items automatically for the topic slot that is active at any time, so that the items in any generated list are diverse, and across many such lists that are generated the desired fairness criteria is also met. The rate at which lists are recomputed can be configured based on the rate at which new items are generated.

Note that the ranking of items in a list needs to be considered as well because there is a rapid drop in the probability of users going further and further down the list to listen to items. Figure \ref{fig:rank} shows the probability of users reaching up to a given rank, as seen from the call logs. We explain next how we incorporate this in our fairness and diversity goals, and define these terms more precisely.



\begin{enumerate}

\item \textbf{Diversity:} We want to ensure that there is sufficient diversity with respect to the aspects covered by the items selected each time a list is generated. We use the diversity sensitive ranking algorithm by Celis, et al \cite{elisa} for this purpose. The algorithm lets us specify an upper-bound at each rank $r$ in the list for the number of items of each aspect that should be allowed to be present in the top $r$ positions in the list. We explain later in the section how these upper-bounds are specified.

\item \textbf{Fairness:} We define the exposure achieved by an aspect as the number of times any item belonging to the aspect is heard by some user. We then build a notion of aspect-level fairness for the exposure achieved by any aspect over a sufficiently long period of time to be in accordance with the specified policies. We work with two fairness policies in this paper: To ensure that each aspect should achieve a certain minimum amount of exposure, or to ensure that all aspects should get an equal amount of exposure. We achieve this fairness guarantee by estimating in advance the desired exposure for any aspect that an item belonging to that aspect should achieve, and then track the amount of exposure actually achieved by each item over time. Whenever a new list of items is generated, we then prioritize the selection of items based on the remaining exposure required for each item. This serves as the primary ranking metric for the diversity algorithm \cite{elisa} used above: Items are ranked based on their required remaining exposure, and the ranking is then adjusted to accommodate the diversity criteria.


\end{enumerate}

This is the algorithm in a nutshell, that for each topic slot, we rank the items based on the remaining exposure they need to achieve, and use the diversity algorithm to adjust this ranking for achieving aspect level diversity as well. This list of items is regenerated at a configurable frequency that can be based on the rate of new item generation for that topic, so that these new items are able to quickly find their way into the lists.

To operationalize this, we next explain three components. First, since we do not have sufficient data per-user to be able to learn the preferences for each user, we identify clusters of users who are similar to one another, and build a classifier that learns the preferences for each cluster. Second, we use the output of the classifier to calculate the desired exposure for each item, for the different fairness policies we implement. We describe these policies and the calculation methods. Third, we describe the algorithm to achieve the diversity and fairness goals.

A few limitations of the current work are listed next, which we aim to address as part of future work. First, the number of users expected to call the platform at any time, and the depth to which users browse the items for a topic, are assumed to be known in advance. Due to budget constraints the social enterprise only permits a certain maximum usage each hour, and hence the hourly traffic on the platform and browsing depths have been found to be quite steady. If however in an unconstrained scenario, the platform traffic volume ends up being more than expected, or the interest in a topic if different from the norm, suitable re-computations can be done to calculate the expected exposure more accurately. Second, we have not addressed the cold-start problem for new users. This can be done by mapping new users to a special cluster for which content is recommended based on a random selection of items in line with the desired fairness policies, and once more data is available about the preferences of the user then they can be mapped to the appropriate cluster. Third, we have also not addressed the cold-start problem for new topics to know in advance the different aspects for the topic on which users may contribute content. We are building a topic modeling step for this purpose, by crawling external data sources such as regional newspapers on a daily basis, so that by the time a new topic becomes popular on the MV platform, we would have accumulated sufficient articles from the mass media and can run topic modeling using tools such as LDA to know which aspects could be likely to surface about the new topic on MV. If needed, each aspect can even further be sub-divided based on varying sentiments of items within the aspect. We leave all these topics for future work.

\subsection{Content recommendation}
We first want to identify clusters of users who have similar preferences as one another. We do this based on indicators for content likeability, and obtain clusters of users with similar tastes in content. Other methods can also be used provided the required data is available. For example, community detection on the underlying social network of relationships between users can be used if the social network information is available \cite{sethicwsm}. Alternately, similarity scores can be calculated for each pair of users based on their shared interests in content, and clustering or matrix factorization approaches can then be applied \cite{collabfiltering}. In our case, we do not have any underlying information about the social network between users, hence we use the second approach.

We begin with mapping each content item to the topic it belongs to. Next, we identify the source of each content item, which can be of two types: created by the content team of MV, or contributed by the users. For each of these (topic, source) pairs, we then develop two indicators: A positive indicator based on how many items for that (topic, source) pair did we notice a positive interaction by the users, and a negative indicator based on how many items saw a negative interaction. Likes, forwards, comments, or a substantially long duration for which an item was heard, are considered positive reactions. Skips within a short duration, and call hang-ups, are considered as negative indicators. We are thus able to build a vector $P_{u}$ for each user $u$, to indicate their positive and negative preference for each (topic, source) pair. A clustering is then done to obtain clusters of users who have similar tastes in content. We expect that this method will translate to some natural demographic clusters, such as for young men who would show a preference for news and jobs related content, women who would show a preference for health and nutrition content, older men who would show a preference for agricultural content, etc.

Next, a topic-specific classifier is developed for each cluster of users, to predict for any item, whether the item should be recommended to the cluster of users or not. A feature vector is developed for each item, consisting of the following parameters:
\begin{enumerate}
\item The aspects to which the item belongs: $A_{ijk} = 1$ if item $i$ belongs to aspect $j$ of topic $k$. This is currently assigned manually by the moderators, but as explained in Figure \ref{fig:content_pipeline} it may be possible to automate this.
\item The rating: $R_i$, which is assigned by the moderators based on their judgment about the quality of item $i$. Moderators assign a rating value between 1..5, where lower values typically signify poor audio quality, and higher values are used to rate based on the actual content in the item, ie. whether it is informative, unique, appears to be useful, etc. In the long run, even user signals such as likes and forwards, or ratings given by community reporters in the field, can be used.
\item The shared context between the contributor of the item and the cluster for which the classifier is being learned: $C_{i}$. This deserves some explanation. As mentioned above, each user is mapped to a cluster based on the content interests of the user. This is also expected to represent an embedding of each user in a social space that the user inhabits in real life. Any content contributed by the user can be expected to reflect this embedding, for example, content contributed by young men is likely to reflect their perspective, and similarly content contributed by women on the same topic is likely to come from a different perspective. Seth, et al verified this hypothesis and introduced the concept of \emph{context} of a user, as representing the social embedding of the user, which manifests itself in any content created by the user \cite{sethwww}. We use this concept in the paper as follows. The preference vector $P_{u(i)}$ is nothing but the context of the user $u$ who contributed item $i$. Similarly, we can calculate the preference vector for the cluster for which the classifier is being learned, as the centroid of the preference vectors of the users belonging to the cluster. We then calculate the shared context between the cluster and the contributor of the item, as the cosine similarity between the two vectors.
\end{enumerate}

The output variable for the item is defined based on the number of positive and negative interactions seen on the item by users of the cluster under consideration. A simple metric of the number of positive interactions minus the number of negative interactions, normalized by the total number of interactions, is used as the output variable. A classifier is then learned to predict whether an item will be liked or not by the users of the cluster. Once such a classifier has been learned for a topic, any chances of a new item being liked by the users in the cluster can be inferred. These items can then be included in the diversity and fairness algorithm that is explained in next section.


\subsection{Content recommendation results}

We next describe the dataset, the offline procedure to identify clusters of users similar to one another, and the content recommendation performance for the classifiers for each cluster of users. 

\subsubsection{Dataset}
We use call logs from a Mobile Vaani deployment in a single district in rural central India, where an independent platform was set up primarily focused on creating awareness and behaviour change for better maternal and child health and nutrition practices \cite{jeevikaictd2019}. Over a period of about 18 months, more than 2 million calls were made by approximately 30,000 callers who listened to over 13,000 items. A description of the data entries in the call logs is shown in Table~\ref{tab:data}. Slots for different topics were maintained on the platform, and typically changed on a monthly frequency. Some examples of these topics included MDD (dietary diversity for mothers), CF (complementary feeding for small children), agriculture, child education and career counseling, local news and updates, etc. Items were mapped to their primary topics, and aspects within the topic to which they belonged. The set of aspects was determined by the project research team, and efforts are currently underway to obtain these aspects automatically using LDA like topic-modeling approaches. Aspects under CF were the following: Basic awareness items carrying information about the importance of complementary feeding of children, in-depth knowledge items carrying detailed information about food groups and food preparation, items discussing common myths and misconceptions, and items about personal experiences and current practices followed by families including reasons why complementary feeding was not pursued activity in their households.

\begin{table}
  \caption{Description of the Mobile Vaani call logs}
  \label{tab:data}
  \begin{tabular}{ccl}
    \toprule
        Data & Description \\
    \midrule
        Cdr\_id & Unique ID of the call \\
        Caller ID & Unique Callerid of the caller \\
        Item ID & Unique ID of the item heard \\
        Contributor & Callerid of caller who recorded the item \\
        Item duration & Duration of the item \\
        Duration heard & Duration of the item heard by the caller \\
        Source & (User/ Studio/ Reporter) Generated Content \\
        Topic & Topic of the item \\
        Aspect & Aspect within the topic of item \\
        Rating & Rating of item given by moderators \\
        Key pressed	& Key pressed by caller while listening to the item \\
  \bottomrule
\end{tabular}
\end{table}

\subsubsection{Obtaining user clusters}

Not all MV users are savvy with using mobile phones, especially rural women users who were the primary target group in this particular project. Many users simply call and listen to whatever audio messages are playing on the IVR, and do not even press buttons to navigate across different items. For the purposes of this paper, we therefore wanted to identify a smaller subset of users who showed a deliberate preference in their content listenership. This was done through a cascading process as follows.

\begin{figure*}[h]
\begin{tabular}{ccc}
    \subcaptionbox{Cumulative distribution of the number of users vs their number of calls, on a log-log scale. The red line indicates the cutoff chosen for the minimum number of calls to have been made by a user to be selected.}{\includegraphics[height = 1.8in]{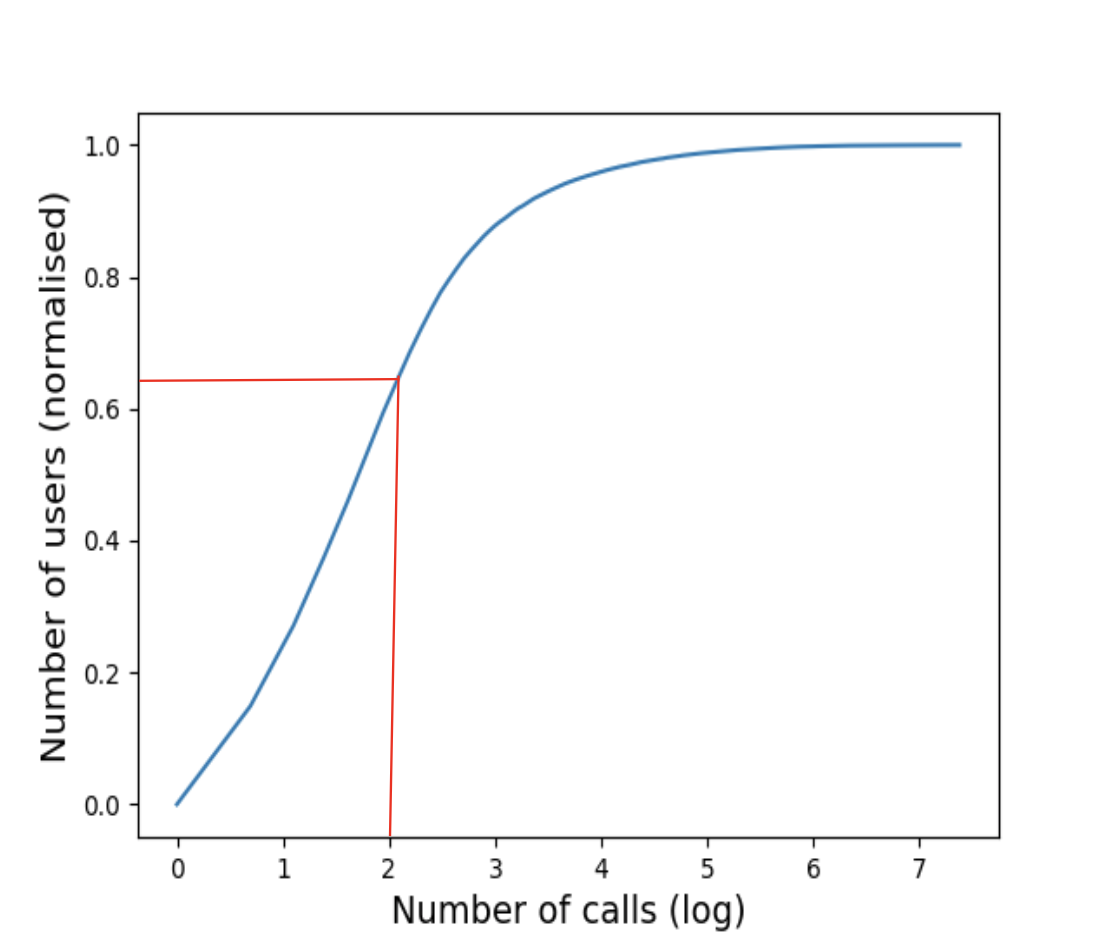}} & \subcaptionbox{Cumulative distribution of keys pressed per second (kps), with kps reported on a negative logarithmic scale. The red line indicates the cutoff chosen for the minimum number of keys to have been pressed per second by a user to be selected as a power user.}{\includegraphics[height = 1.8in]{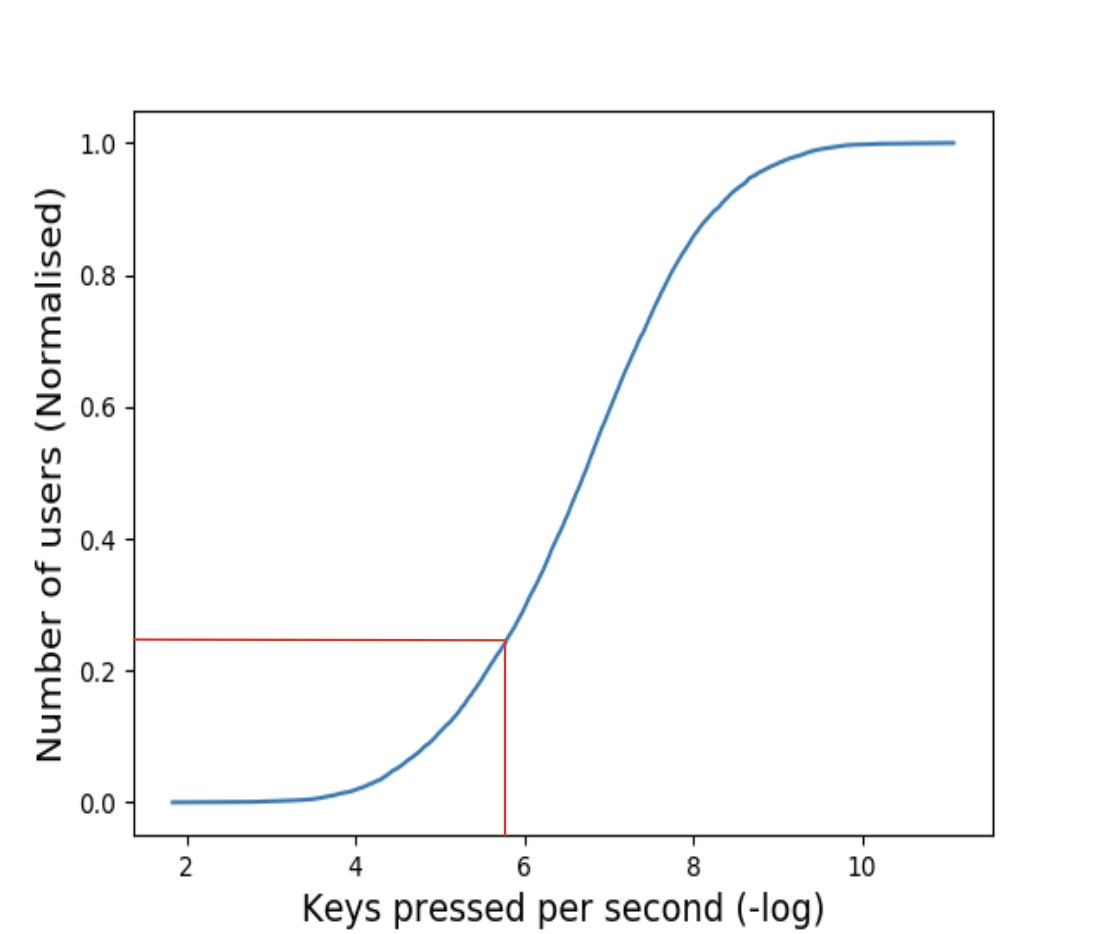}} & \subcaptionbox{Cumulative distribution of the number of users vs the KL divergence scores scaled by a factor of 10 on a negative logarithmic scale. The red line indicates the cutoff chosen for the minimum divergence score.}{\includegraphics[height = 1.8in]{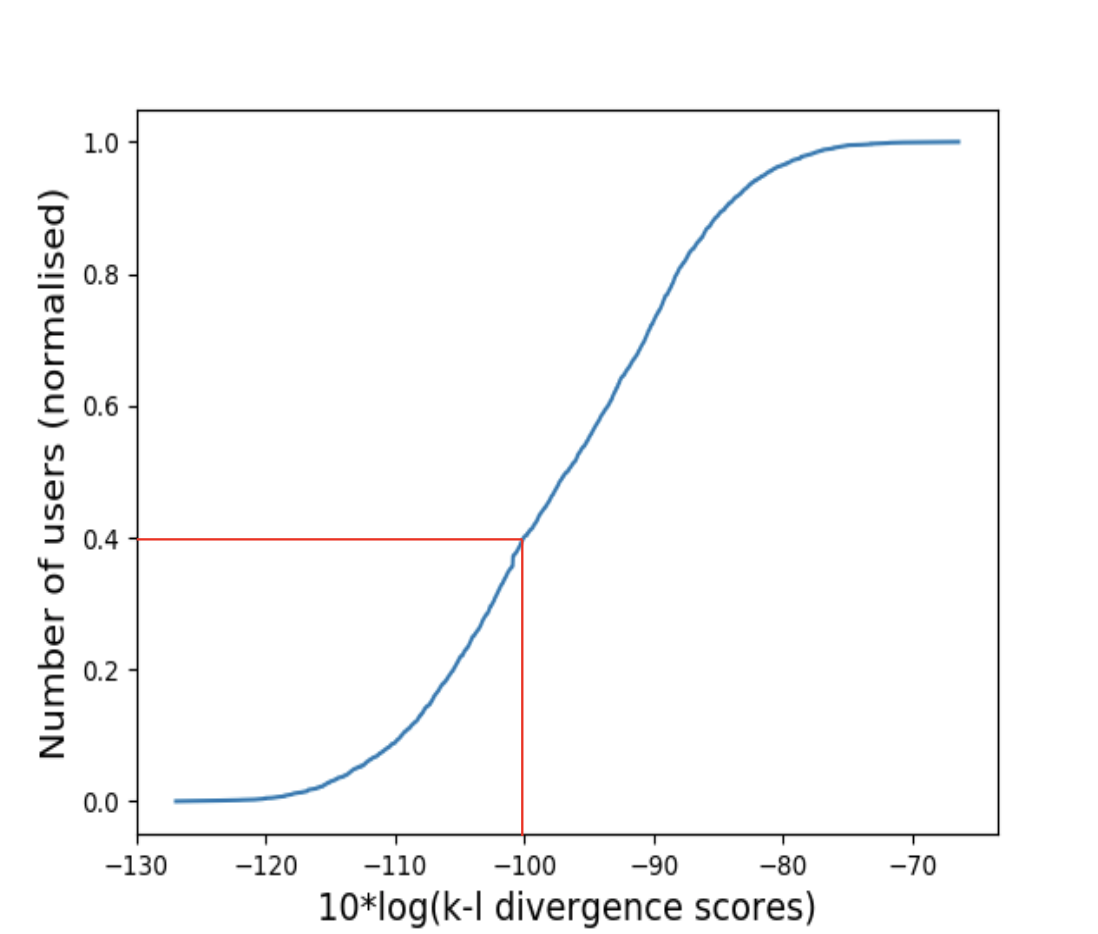}}
\end{tabular}
\caption{Identification of users for evaluation of the algorithms}
\label{fig:pue}
\end{figure*}

We first identified frequent callers who had called more than seven times over the duration of the dataset. A distribution of users according to their number of calls is shown in Figure \ref{fig:pue}a, the chosen threshold gave us about 35\% users who had called reasonably often. For these users, we then looked at whether they were savvy users and pressed keys to navigate. We estimate the key-pressing frequency over the entire lifetime of a user, and interpret it as a signal of whether users actively makes choices in the content they like to listen to, under the assumption that they will not like all the content they get to hear. Figure \ref{fig:pue}b shows the key presses per second for each user, over all the calls made by them. A threshold was chosen of approximately one key pressed every four minutes, and left us with 25\% of the callers.

Out of these, we further wanted to select users who were substantially different in their listening patterns from the average. This is because users typically take time to learn how to use the system, and we wanted to eliminate data from initial periods where they may have been listening to whatever content was presented to them, without exercising their preference. We chose to do this by observing how different was the preference vector for a user from the global preference vector across all the users. Other methods could also have been used, for example for each user to only consider data after their key-press frequency had increased. Figure \ref{fig:pue}c shows the distribution of KL divergence, calculated between the preference vector for a user and the global preference vector across all users. A threshold was used to select 60\% of the users, eventually leaving us with 2078 users for our evaluation.

After having filtered the users to a smaller set for evaluation, we proceeded to cluster them based on their preferences shown to items belonging to different (source, topic) pairs. The preference score for a user towards a particular (source, topic) pair was calculated as follows:
\begin{equation}
    P\_score = \frac{|positive\ interaction\ items| - |negative\ interaction\ items|}{|items\ heard|}
\end{equation}

We chose this metric rather than retaining positive and negative interactions separately due to data sparsity issues that would lead to poor clusters, despite using dimensionality reduction techniques like PCA. As described earlier, a positive interaction is interpreted as a liking or forwarding or commenting action taken on an item, or having heard more than 45\% of the item. Similarly, a negative interaction is interpreted as an item skip or call hangup before 45\% of the item was heard. An additional binary indicator column was added to distinguish two cases of (a) a user not having heard any item of a particular (source, topic) pair, and (b) having heard items but the positive and negative interactions having canceled each other out.



With the preference vectors thus formed containing both continuous and categorical variables, we ran the k-prototype algorithm for clustering. The optimal value of $k$ was chosen to be $5$ based on the elbow-curve that plots the cost (intra-cluster distance divided by the inter-cluster distance) for different values of $k$.



\subsubsection{Classification of items}
We next build topic-specific classifiers for each cluster of users, to decide whether or not to recommend an item to the users of the cluster. For the experiments shown in the paper, we only do this for the MDD (maternal dietary diversity) topic, results for other topics are similar and shown in \cite{sm}. MDD had five aspects. We built a feature vector for each item consisting of the moderator-assigned rating to the item, indicators variables for each aspect to indicate whether the item belongs to an aspect or not, and the shared context between the contributor of the item and the cluster's centroid. The output variable for whether to recommend the item to the cluster or not was assigned by calculating an average preference score similar to $P_{score}$ in equation 1, as the difference between the number of users who positively interacted on the item and the number of users who negatively interacted, normalized by the total number of users who interacted with the item. We ignore the items that were not been heard by any user in the the cluster. We then used a \textit{random forest classifier}, with SMOTE (Synthetic Minority Over-sampling Technique) \cite{smote} to address the class imbalance. The results of the classifiers for each cluster are shown in Table \ref{tab:forest}. There is clearly much scope for improvement, the models are not doing well for all clusters, but we continue with it since our primary goal is to evaluate the fairness and diversity algorithms. The search for better content recommendation models can be pursued independently. 

In the next section, we evaluate these algorithms for only the first two clusters where a reasonable accuracy is obtained by the content recommendation classifiers. Other clusters for which the classifiers are not working well, can be treated as a cold-start scenario where enough information is not available for new users.

\begin{table}
  \caption{Validation accuracy for MDD random forests classifier}
  \label{tab:forest}
  \begin{tabular}{ccc}
    \toprule
        Cluster & Validation accuracy(\%)\\
    \midrule
        1 & 73.8\\
        2 & 75\\
        3 & 63.3\\
        4 & 56\\
        5 & 43.6\\
  \bottomrule
\end{tabular}
\end{table}

\section{Fairness and diversity algorithm}

The topic-specific cluster-specific classifiers described in the previous section are able to convey for each item belonging to the given topic, whether it will be liked by users belonging to the given cluster or not. We can thus obtain the set of items $B_{jk}$ belonging to aspect $j$ for topic $k$, that will be liked by users belonging to the cluster. Across all the aspects ($\bigcup_{j}B_{jk}$), we are thus able to obtain $\beta_{jk}$ as the fraction of items for aspect $j$ that are liked by the users.
Considering each item of an aspect as equally important, we can now begin to define our fairness policies. The first policy where we want to ensure that a certain minimum amount of exposure is given to each aspect, is nothing other than ensuring that $\beta_{jk}$ is more than the desired minimum amount for all aspects. If any aspect is under-represented, then items classified negatively can be pulled in, overriding the user preferences. Alternately, items from under-represented aspects can be duplicated within the aspect, effectively increasing the number of items in this aspect to offset the under-representation. Yet another method can be to first calculate the desired exposure for each aspect, and then divide that exposure equally among all the positively classified items for that aspect; we use this method in our current implementation. The second fairness policy where we want to ensure that all aspects are given an equal exposure, can be similarly implemented as well.

\begin{figure}[!ht]
    \centering
    \begin{subfigure}{1\columnwidth}
        \centering
        \includegraphics[height=1.7in, width=0.75\columnwidth]{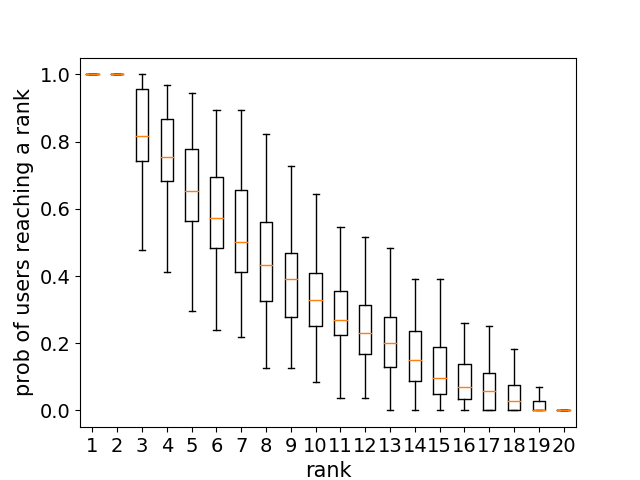}
        \caption{Probability of users reaching a rank}
        \label{fig:rank}
    \end{subfigure}
    \begin{subfigure}{1\columnwidth}
        \centering
        \includegraphics[height=1.7in, width=0.75\columnwidth]{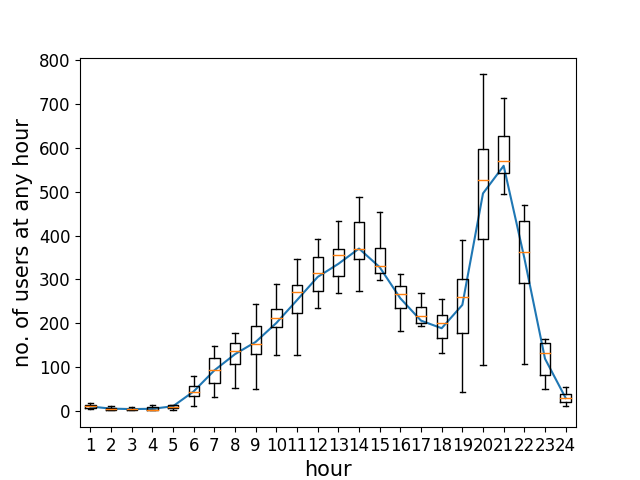}
        \caption{Mean number of users who call at different hours in a day}
        \label{fig:users_per_hour}
    \end{subfigure}
    \caption{Distribution of users across ranks in a list and different hours in a day}
    \label{pic:uninuo}
\end{figure}

Once the final set of items $B_{jk}$ has been prepared in accordance with the fairness policies, the next step is to calculate the desired exposure for each item. This is done as follows. Since we assume an invariant in the number of users calling the platform at any given time (Figure \ref{fig:users_per_hour}), and an invariant in the probability of users browsing items to greater depths in the IVR lists (Figure \ref{fig:rank}), we can calculate the total inventory for exposure available for a given topic, based on the slots identified for the topic by the content operations team. The desired exposure for any item is then nothing but this total amount of inventory distributed equally among all the items, or first distributed among all the aspects based on the $\beta_{jk}$ proportions and then divided equally within an aspect among all the items belonging to the aspect. We are thus able to calculate the desired exposure $D_{ijk}$ for each item $i$. A memory variable $E_{ijk}$ is additionally maintained for each item to keep track of the exposure already attained by the item at any point in time. Whenever generating a new ranked list of items, the ranks are calculated based on the difference $D_{ijk}$ - $E_{ijk}$, so that items that have received less exposure until then are ranked higher than items that have obtained more exposure.

For each of the two fairness policies, we also build an additional variant. We explained above that the total inventory was divided equally among all the $B_{jk}$ items, or first divided among the aspects in proportion to $\beta_{jk}$ and then divided equally among the items in that aspect. This method ensures equality in the exposure given to the items. In a variant, we divide the inventory in proportion to the ratings $R_i$ given to the items by the moderators. This can serve as another editorial control mechanism, to augment the fairness policies that can be specified editorially.

Given the time slots assigned to a topic over which the total inventory is calculated, we divide them into fractional slices of short duration for which a new list of ranked items is generated. A draft ranked list is first generated based on the additional exposure needed by different items, ie. $D_{ijk}$ - $E_{ijk}$, as explained above. This list is then adjusted to ensure short-term diversity, based on the algorithm by Celis, et al \cite{elisa}. This algorithm operates as follows. It requires an input of $m$ items sorted in decreasing order of their utility, which in our case is the additional exposure needed by an item. It then outputs a list of $n$ << $m$ items, obeying the diversity constraints. This is done through a greedy approach proven to be correct.

To satisfy the short-term diversity criteria, for a given list of items and their corresponding utility values (in our case, the remaining exposure), the algorithm iteratively constructs a ranking $\pi$ : [$n$] -> [$m$] (i.e. $\pi$(j) is the item ranked at position $j$, for all $j$ $\in$ [$n$]). 

\begin{algorithm}
  \caption{Short-term diversity}\label{sd}
  \begin{algorithmic}[1]
    \Procedure{Short-term diversity}{$items,utility$}
      \For{$j$ = $1$ to $n$}
        \State Let $i$ $\in$ [$m$] be the smallest index of an item which has not yet been picked and can be added at position $j$ without violating any constraint. If there is no such $i$, output the smallest index item which has not yet been picked.
        \State Set $\pi$(j) = $i$ 
      \EndFor
      \State \textbf{return}$\pi$
    \EndProcedure
  \end{algorithmic}
\end{algorithm}


The upper-bound constraint $U_{jp}$, $1<=p<=n$ can be defined as the number of items of aspect $j$ that are allowed to appear in the top $p$ positions of the final ranking. We choose the following constraint in our current work. 

\begin{equation}
    U_{jp}=p*\beta_{jk} 
\end{equation}

That is, the top-$p$ items in a list will have the same distribution as the proportion of exposure desired for different aspects. 

The total time period is divided into slots with new lists generated for each slot. To satisfy the long-term fairness criteria, we call the Short-term\_diversity algorithm for each slot after updating the utility as the remaining exposure for each item. The algorithm operates as follows, taking the set of positively classified items as an input for each cluster:


\begin{algorithm}
  \caption{Longterm fairness}\label{lf}
  \begin{algorithmic}[1]
    \Procedure{longterm fairness}{$items$}
    \State utility$_{ijk} = D_{ijk}$
      \For{$s$ = $1$ to $\#slots$}
        \State $\pi$ = \textproc{Short-term diversity}(\textit{items, utility})
        \State Output $\pi$
        \State utility$_{ijk} = D_{ijk} - E_{ijk}$
      \EndFor
    \EndProcedure
  \end{algorithmic}
\end{algorithm}

This method is thus able to ensure both short term diversity whenever a new ranked list is generated, and long term fairness based on the specified policies. Other constraints for diversity can also be defined, for example, to ensure that at least one item for a given important aspect always features in a list, or consecutive items always belong to different aspects, etc. 


\section{Experiments}

We next evaluate our algorithms to compare the diversity and fairness properties achieved in three kinds of setups:
\begin{enumerate}
    \item Manual moderation as is done currently, where items are ranked entirely based on the discretion of the content operations team
    \item Recommendations in line with user preferences as inferred by the content recommendation step
    \item Recommendations which adjust the above to incorporate diversity and fairness constraints. We evaluate four kinds of fairness policies:
    \begin{enumerate}
        \item A certain minimum exposure guaranteed to each aspect, with equal exposure to items within each aspect
        \item Minimum exposure to each aspect, with per-item exposure proportional to the moderator assigned rating for the items
        \item Equal exposure to each aspect, and equal exposure to items within each aspect
        \item Equal exposure to each aspect, and per-item exposure proportional to the rating of the items
    \end{enumerate}
\end{enumerate}

We use a random selection model for a baseline, which chooses the next item for each rank uniformly randomly from among the items that have not been selected so far.

\subsection{Comparison of models}
We run each of the above models by simulating the call logs, ie. we replay the given workload and do not generate an artificial model-driven workload. The call logs provide us with the set of users who called during each MDD slot. To simulate the manual moderation model, we simply obtain the set of items that were actually heard by the users. For each of the other models, we generate the list of items based on the different policies. We generate a new list of 10 items every hour, given that the rate of new item generation seen in the MDD topic is approximately 0.4 items per hour, hence new items are soon able to find their way into the list. To calculate the desired exposure for each item, we use a future time horizon of 100 hours over which to ensure long term fairness, which effectively translates to one month of exposure on a slot-allocation of three hours per day for MDD. In actual practice, these values would be configured based on the traffic volumes.


We next present a comparison of the different models based on the following: Long term fairness, short term diversity, and the deviation in the exposure given to various items by our models from the model that predicts only based on the user preferences without taking any fairness and diversity constraints into account. In the models 3a and 3b which provide a minimum exposure to each aspect, we chose the minimum exposure to be 5\% of the total exposure.




\begin{figure*}[h]
\begin{tabular}{cccc}
    \subcaptionbox{Distribution of exposure achieved by various aspects, for different models}{\includegraphics[height = 1.8in]{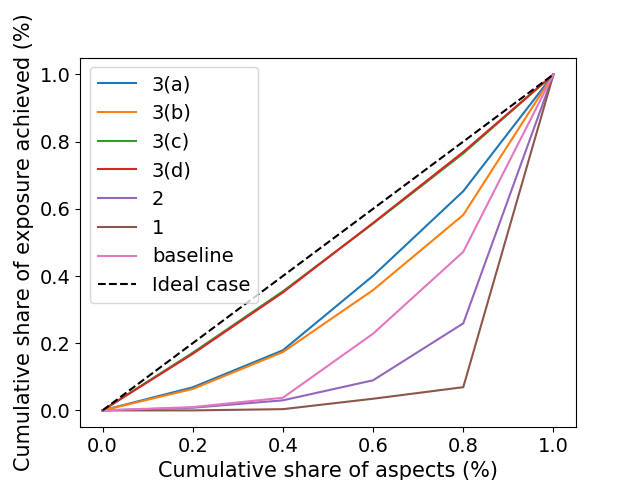}} & \subcaptionbox{Distribution of HHI over all list generations}{\includegraphics[height = 1.65in]{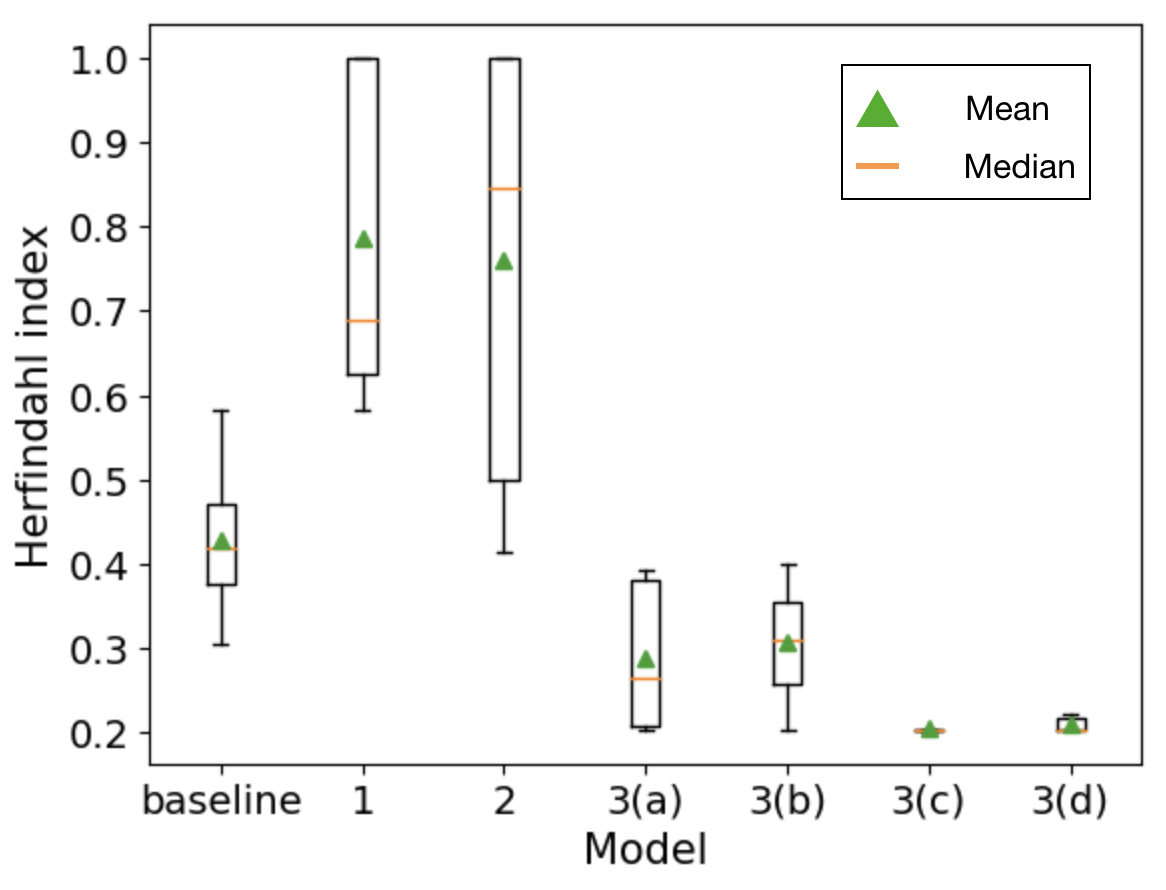}} & \subcaptionbox{Deviation from perfect user satisfaction: Normalized RMSEs for models 3a..3d}{\includegraphics[height = 1.8in]{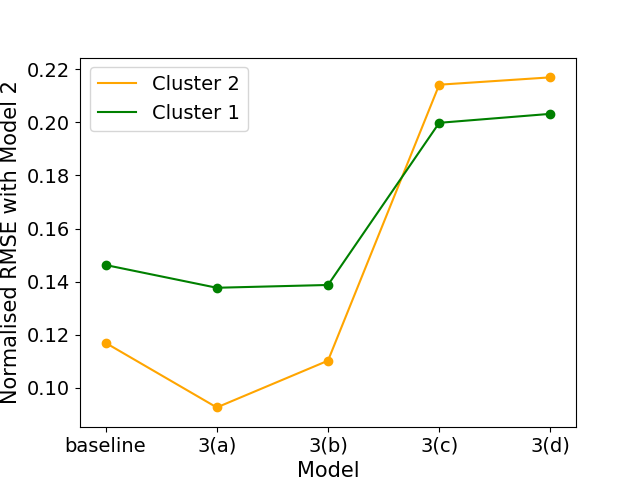}}
\end{tabular}
\caption{Comparison of models. Figures a and b show combined results for users of both the clusters}
\label{fig:com}
\end{figure*}

\subsubsection{Fairness}
We plot the Lorenz curves (Figure \ref{fig:com}) and calculate the Gini coefficients (Table \ref{tab:gini}) to evaluate the fairness achieved by different models in terms of the exposure given to the various aspects. Models close to the X=Y line, and those with a low Gini coefficient, are more fair. The two models for manual moderation (model 1) and presenting items only according to the user preferences (model 2) show the worst fairness, poorer than the baseline. Among the other models, as expected the ones forcing an equal exposure to all the aspects (models 3c and 3d) perform close to perfectly. The others ensuring a minimum exposure to each aspect (models 3a and 3b) do not perform as well as 3c and 3d, but better than the baseline.



\begin{table*}
  \caption{Gini coefficients for different recommendation models, indicating the fairness achieved by various aspects}
  \label{tab:gini}
  \begin{tabular}{ccc}
    \toprule
        Model no. & Model Description & Gini coeff value \\
    \midrule
        0 & Random selection (Baseline) & 0.500\\
        1 & Manual moderation & 0.756 \\
        2 & User preferences & 0.645 \\
        3a & Aspect(min guarantee), Item(equal exposure) & 0.279 \\
        3b & Aspect(min guarantee), Item(exposure proportional to rating) & 0.328 \\
        3c & Aspect(equal exposure), Item(equal exposure) & 0.061 \\
        3d & Aspect(equal exposure), Item(exposure proportional to rating) & 0.061 \\
  \bottomrule
\end{tabular}
\end{table*}

\subsubsection{Diversity}
We evaluate the models on diversity by calculating the Herfindahl-Hirschman (HHI) index for each list, on the diversity of aspects in the list \cite{hhi}. HHI is calculated as the sum of squares of the shares of each element, and is often used as a measure of market concentration in economics. Higher HHI values denote higher concentration, ie. monopoly power. Figure \ref{fig:com} shows the box-plot for HHI values across all instances of the list generations. Clearly all of our proposed models (3(a) - 3(d)) offer a significant improvement in diversity due to the constraints imposed by the ranking algorithm. In the case of models 1 and 2, no attention was paid on whether the items in a list belonged to the same aspects or not, and resulted in HHI scores even higher than the baseline.


\subsubsection{Deviation from user satisfaction}
Model 2 recommends items in accordance with the user preferences. We took this as the ideal exposure that should be achieved by the items, and calculated the deviation in exposure given to each item by our models. Figure \ref{fig:com}c shows the mean normalized RMSE scores for the four models separately for the two clusters. In both cases, we can see that models 3a and 3b allocate exposure closest to the user preferences because they only impose a small deviation to ensure some minimum exposure to all aspects, hence the normalized RMSEs for these models is the lowest, and does not cause significant deviation from ideal user satisfaction. Models 3c and 3d provide an equal exposure to all aspects, and hence deviate more than the ideal user satisfaction, even more than the baseline.


\begin{figure*}[htbp]
\begin{tabular}{c|c|c}
    \subcaptionbox{Model 1 with baseline inset}{\includegraphics[height = 1.8in]{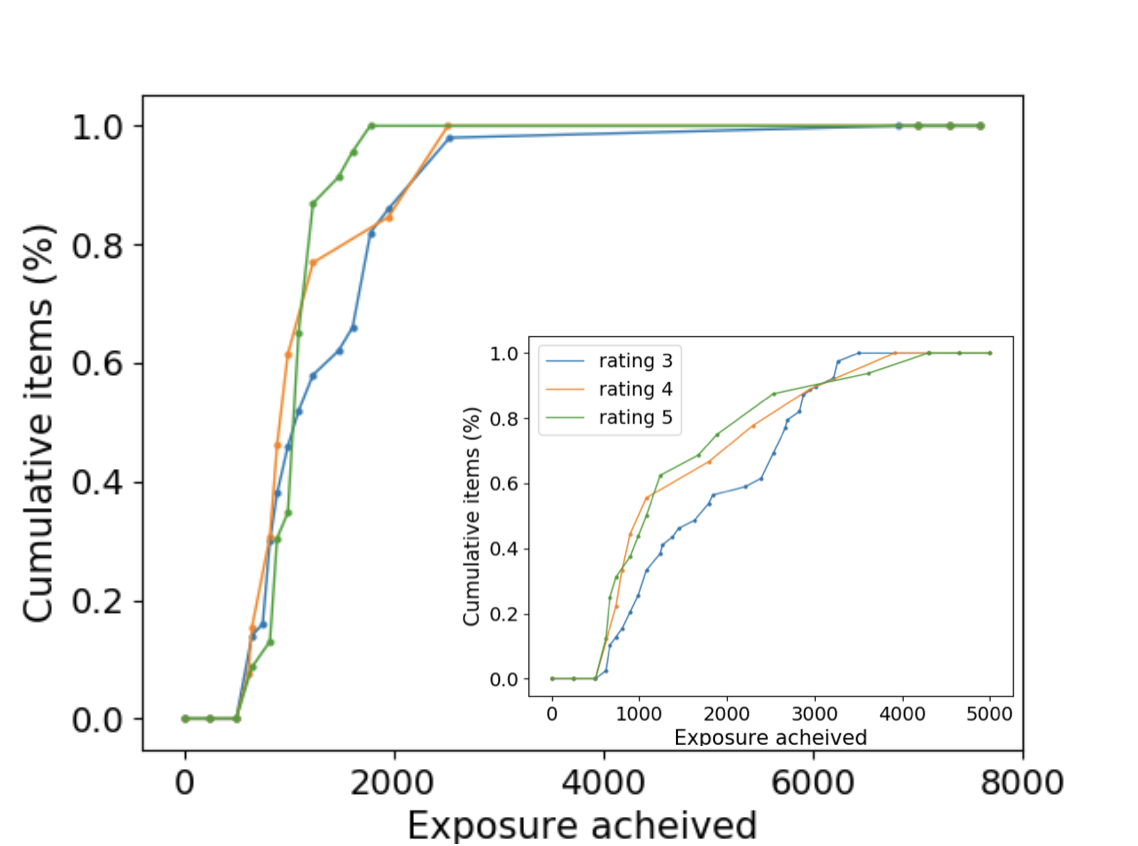}} & \subcaptionbox{Model 3(a)}{\includegraphics[height = 1.8in]{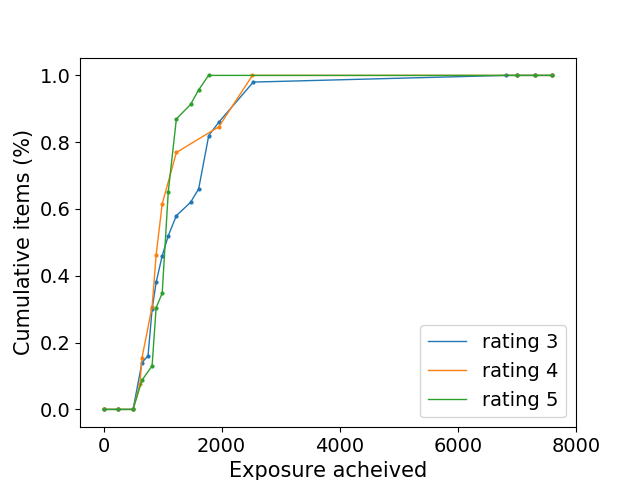}} & \subcaptionbox{Model 3(c)}{\includegraphics[height = 1.8in]{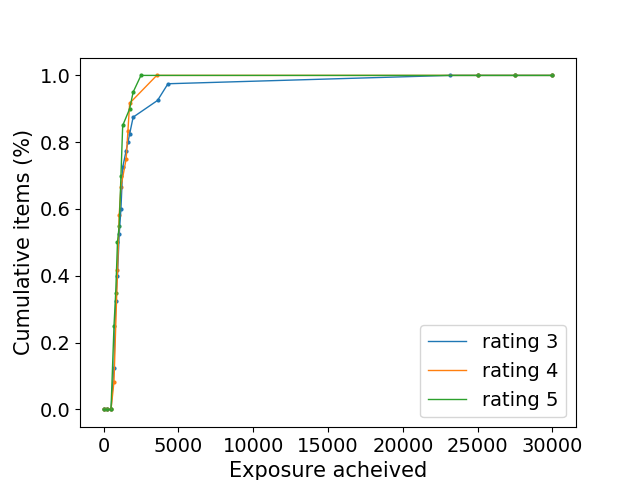}} \\
    \subcaptionbox{Model 2}{\includegraphics[height = 1.8in]{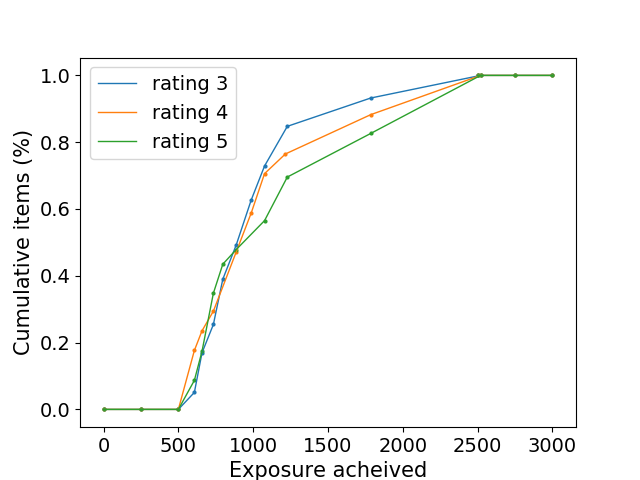}} & \subcaptionbox{Model 3(b)}{\includegraphics[height = 1.8in]{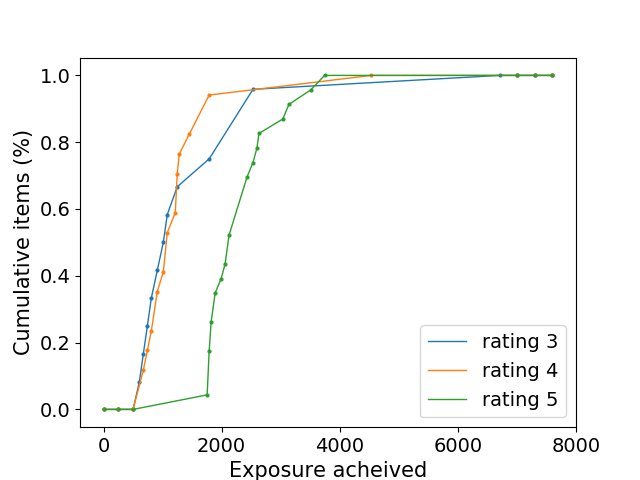}} & \subcaptionbox{Model 3(d)}{\includegraphics[height = 1.8in]{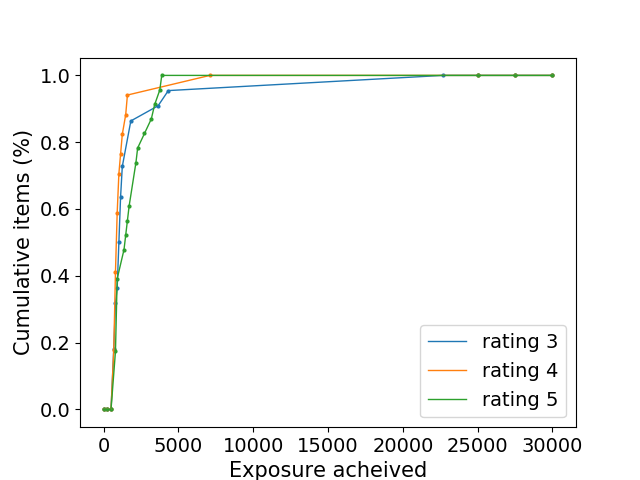}}
\end{tabular}
\caption{Cumulative distribution of exposure achieved by the items for the ratings 3,4 and 5}
\label{fig:rating}
\end{figure*}

\subsubsection{Exposure to items of different ratings}
We also check how well the editorial policies for models 3b and 3d are obeyed, to give exposure to items in proportion to their moderator-assigned ratings. Figure \ref{fig:rating} shows the cumulative distribution for exposure achieved by the items for ratings 3, 4, and 5. Model 1 for manual moderation clearly gives a strong preference for rating-5 items, indicating that the content moderators tend to keep highly rated items at the top in the IVR lists. Shown for comparison is the baseline model, which ended up providing exposure to ratings in the order of their frequency, ie. most to rating 3 and least to rating 5. Model 2 which only looks at user preferences, does not distinguish between the ratings, possibly also indicating that the user preferences do not necessarily correlate with what the moderators may feel are high quality items. Model 3b is able to give better exposure to higher rated items than model 3a. Model 3d is similarly able to give more exposure to higher rated items than model 3c, although the exposure seems to be imbalanced favouring a few items only. This happened because in models 3c and 3d, all aspects are given equal exposure, but some aspects had just a few items and that too less rated items, which therefore ended up getting much more exposure than would have been desirable. Overall, the models seem to be working as expected, and with more data in terms of larger volumes of items and more users, we can expect to see the different fairness policies being able to achieve their objectives.

\section{Discussion and conclusions}

In this paper we have presented a content recommendation and ranking model for a participatory media platform that is able to provide certain fairness and diversity guarantees based on editorial policies that can be specified by the platform managers. This can be important to cultivate a culture of respect for diversity among the users, and avoid filter bubbles where users are only presented with views they prefer to see and hear.

There are several improvements that can be made to the model, as have been mentioned in the text above, such as to build methods to handle the cold-start problem for new users, and topic modeling for new topics. The ranking algorithms can also be extended in an interesting manner. The current framework allows the specification of aspect-level exposure for fairness policies, and within each aspect it provides an equal exposure to all items or in proportion to the moderator-assigned ratings given to the items. This can be improved so that items are given exposure in proportion to live-feedback obtained about them, based on the positive and negative interactions by other users on the items. Further improvements can be made to provide an equal opportunity for new items to become popular: This can be done by having two levels of desired exposure, initially to give a minimum exposure to each new item to be able to observe how users react to it, and then to give an exposure based on the positive and negative feedback obtained on it.


Participatory media systems also face a serious problem of content credibility, as is known from recent episodes about fake news where factually incorrect information was published on social media platforms and led to unfortunate outcomes \cite{indiawhatsapp, trumpelections, debroytwitter}. Since the MV platform is moderated, issues of objective credibility such as publication of fake or factually incorrect news have been successfully addressed. A large team of community reporters in the field are contacted by the moderators to verify any alarming news or allegations recorded on the platform before this information is published. Efforts are underway to reduce verification delays by decentralizing the moderation step to the community reporters. Automated steps can also be used to flag seemingly suspicious content, that might be seeing strong engagement in some user clusters but not others, as suggested in \cite{sethwww}. Some times information may also raise subjective concerns of credibility, ie. the information may not be factually incorrect but could represent views that seem less-credible to some users \cite{sethwww}. Since our goal however is to provide diverse viewpoints to users, we do not want to remove such controversial opinions and would rather extend our model to incorporate them. This can be done by increasing the number of classes where currently we have only considered aspects about a topic, but we can easily extend by adding a dimension of sentiment to each aspect, for example.

We feel that this space of content recommendation and ranking in participatory media environments can be instrumental in building a public sphere with a progressive culture of deliberation, as what used to exist in physical environments at one time \cite{habermas}. Participatory media has always been envisioned to offset the biases in mass media \cite{schefeule} and become a voice of the people \cite{liberationtech}, but the open design of platforms like Facebook and Twitter has been argued to not be suited for this purpose because of a lack of agency given to the users to shape the platform use in responsible ways \cite{responsibleoutcomes}. Other platforms like Reddit and Mobile Vaani may be more empowering in their fundamental design, and editorial algorithms can help amplify their capabilities to encourage more discussion and diversity to build a vibrant public sphere.



\bibliographystyle{ACM-Reference-Format}
\bibliography{main}

\end{document}